\documentclass[review]{elsarticle}%
\usepackage{lineno,hyperref}
\usepackage{amsmath}
\usepackage{amsfonts}
\usepackage{amssymb}
\usepackage{graphicx}%
\modulolinenumbers[5]
\journal{Journal of \LaTeX\ Templates}
\begin{document}

\begin{frontmatter}
\title{Anomalies in weak decays of H-like ions}
\author{Francesco Giacosa}
\address{Institute of Physics, Jan Kochanowski University, ul.\ Swietokrzyska
15, 25-406 Kielce, Poland}
\author{Giuseppe Pagliara}
\address{Dip.~di Fisica e Scienze della Terra dell'Universit\`a di Ferrara and INFN
Sez.~di Ferrara, Via Saragat 1, I-44100 Ferrara, Italy}
\begin{abstract}
We investigate the emergence of oscillations in the decay law of unstable systems. We discuss
in particular the case of the so-called GSI anomaly seen in the electron capture decays
of H-like ions and prove that such oscillations cannot be explained by neutrino oscillations.
We then discuss how such anomalies could be intimately related to the decay law of unstable
systems in the case in which their spectral function deviates from a Breit-Wigner shape.
\end{abstract}
\end{frontmatter}

\linenumbers


\section{Introduction}

The decay of the H-like ions $^{140}$Pr and $^{142}$Pm via electron capture
was measured at the GSI storage ring \cite{gsi}, where some peculiar
oscillations in the decay law were found. In some works \cite{ivanov} these
anomalies were linked to the phenomenon of neutrino oscillations. However,
this interpretation has been also heavily disputed in other articles, see for
instance \cite{merle}.

In this work we first confirm, by using the Lee-Hamiltonian formalism
\cite{lee} for the study of decays \cite{facchiprl,duecan}, that neutrino
oscillations cannot generate time modulations in the experimental set-up of
the GSI experiment. Then, we discuss the emergence of a non-exponential decay
law as a consequence of deviations from the Breit-Wigner energy distribution.
Indeed, a decay law which presents clear oscillations was measured in the
decay thorough tunneling of sodium atoms in an accelerated optical potential
\cite{reizen}, thus showing that the short-time deviations from the decay law
are an experimental fact.

\section{Neutrino oscillations: why they cannot generate time modulations}

The process under study is schematically given by $M\rightarrow D+\nu_{e}$
,where $M$ stays for the $H$-like mother state (such as $^{140}$ Pr) and $D$
for the daughter nucleus state (such as $^{140}$Ce ). Because of neutrino
mixing ($\nu_{e}=\cos\theta\nu_{1}+\sin\theta\nu_{2}$) one obtains the decay
into two channels: $M\rightarrow D+\nu_{1}$ and $M\rightarrow D+\nu_{2}$. The
energy-momentum conservation implies that (in the first channel)
$p=q_{1}+k_{1},$ out of which (in the reference frame of the mother)
$M_{M}^{2}+m_{\nu_{1}}^{2}-2M_{M}E_{\nu_{1}}=M_{D}^{2}$ . A similar expression
holds in the second channel. Then, the energy difference is given by $\Delta
E_{\nu}=E_{\nu_{2}}-E_{\nu_{1}}=\left(  m_{\nu_{2}}^{2}-m_{\nu_{1}}%
^{2}\right)  /2M_{M}$. Now, if the decay amplitude \textit{would} have an
expression of the form
\begin{equation}
A\propto\cos\theta e^{-iE_{\nu_{1}}t}+\sin\theta e^{-iE_{\nu_{2}}t}\text{ ,}%
\end{equation}
then the square amplitude \textit{would} read%
\begin{equation}
\left\vert A\right\vert ^{2}\propto1+2\cos\theta\sin\theta\cos\left(  \Delta
E_{\nu}t\right)  \text{ ,}%
\end{equation}
out of which oscillations with period $T=2\pi/\Delta E_{\nu}$ emerge.
Obviously, this is not a derivation of an oscillation formula. Actually, a
straightforward calculation of the survival amplitude $a(t)$, as shown in the
following, does \textbf{not} lead to an oscillatory amplitude of this kind.
Anyway, it is suggestive that, if we use the present value for the mass
difference $\Delta m_{21}^{2}=7.6\times10^{-5}$eV$^{2}$ \cite{Abe:2010hy}, we
obtain (by setting $M_{M}=132$ GeV) $T=2\pi/\Delta E_{\nu}=4\pi M_{M}/\Delta
m_{21}^{2}\simeq14$ s, which is \textit{remarkably }close to the measured
oscillation period $T_{measured}\simeq8$ s. This is the reason why neutrino
oscillation has been considered appealing, even if it cannot hold.

We now turn to the Lee-Hamiltonian formalism to describe decays \cite{lee}.
This formalism is equivalent to QFT at one-loop \cite{duecan,lupo,zenoqft} (in
most cases a good approximation \cite{schneitzer}), hence the correct
formalism to describe decays. The basis of states that we consider is given by
$\left\{  \left\vert M\right\rangle ,\left\vert D(\mathbf{k}),\nu
_{1}(-\mathbf{k})\right\rangle ,\left\vert D(\mathbf{k}),\nu_{2}%
(-\mathbf{k})\right\rangle \right\}  ,$ where again $M\equiv$ mother (in the
rest frame) and $D\equiv$ daughter. The Hamiltonian $H=H_{0}+H_{1}$ reads:%
\begin{align}
H_{0} &  =M_{0}\left\vert M\right\rangle \left\langle M\right\vert
+\sum_{i=1,2}\int d\mathbf{k}\omega_{i}(\mathbf{k})\left\vert D(\mathbf{k}%
),\nu_{i}(-\mathbf{k})\right\rangle \left\langle D(\mathbf{k}),\nu
_{i}(-\mathbf{k})\right\vert \text{ , }\nonumber\\
H_{1} &  =\sum_{i=1,2}\int d\mathbf{k}\frac{g_{i}f_{i}(\mathbf{k})}%
{(2\pi)^{3/2}}\left(  \left\vert M\right\rangle \left\langle D(\mathbf{k}%
),\nu_{i}(-\mathbf{k})\right\vert +\text{h.c.}\right)  \text{ .}%
\end{align}
The state $\left\vert D(\mathbf{k}),\nu_{1}(-\mathbf{k})\right\rangle $
($\left\vert D(\mathbf{k}),\nu_{2}(-\mathbf{k})\right\rangle $) represents a
two-particle state, a $D$ with momentum $\mathbf{k}$ and a neutrino with
$-\mathbf{k}.$ The energies read $\omega_{i}(\mathbf{k})=\sqrt{\mathbf{k}%
^{2}+M_{D}^{2}}+\sqrt{\mathbf{k}^{2}+m_{\nu_{i}}^{2}},$ $g_{i}$ are coupling
constants and $f_{i}(k)$ are form factors. In the exponential limit,
$f_{i}(k)=1$, the time evolution reads:%
\begin{equation}
e^{-iHt}\left\vert M\right\rangle =e^{-i(M_{M}-i\Gamma/2)t}\left\vert
M\right\rangle +\sum_{i=1,2}\int d\mathbf{k}b_{i}(\mathbf{k},t)e^{-i\omega
_{i}(\mathbf{k})t}\left\vert D(\mathbf{k}),\nu_{i}(-\mathbf{k})\right\rangle
\end{equation}
where $b_{i}(\mathbf{k},t)=\frac{g_{i}}{(2\pi)^{3/2}}\frac{e^{-i\omega
_{i}(\mathbf{k})t}-e^{-i(M_{0}-i\Gamma/2)t}}{\omega_{i}(\mathbf{k}%
)-M_{0}+i\Gamma/2},$ see e.g. \cite{giacosapra}. The survival probability
amplitude of the state $\left\vert M\right\rangle $ is also in this case the
usual exponential form $a(t)=\left\langle S\right\vert e^{-iHt}\left\vert
S\right\rangle =e^{-i(M_{0}-i\Gamma/2)t}$, hence the survival probability is
$p(t)=e^{-\Gamma t}$ (with $\Gamma=\Gamma_{1}+\Gamma_{2}$ and $\Gamma
_{i}=g_{i}^{2}$). Alternatively, one can calculate the probability that the
state has decayed between $(0,t),$ which is given by
\begin{equation}
w(t)=\int d\mathbf{k}\left[  \left\vert b_{1}(\mathbf{k},t)\right\vert
^{2}+\left\vert b_{2}(\mathbf{k},t)\right\vert ^{2}\right]  =1-e^{-\Gamma
t}=1-p(t)\text{ ,}%
\end{equation}
where clearly no oscillations exist. If, instead, we evaluate the probability
to measure the final state in a combination corresponding to a neutrino
$\nu_{e},$
\begin{equation}
\left\vert F_{e}(\mathbf{k})\right\rangle =\cos\theta\left\vert D(\mathbf{k}%
),\nu_{1}(-\mathbf{k})\right\rangle +\sin\theta\left\vert D(\mathbf{k}%
),\nu_{1}(-\mathbf{k})\right\rangle
\end{equation}
(for whatever value of $\mathbf{k}$), we find: $w_{\nu_{e}}=\int
d\mathbf{k}\left[  \left\vert \cos\theta b_{1}(\mathbf{k},t)+\sin\theta
b_{2}(\mathbf{k},t)\right\vert ^{2}\right]  .$ In general, $w_{\nu_{e}}(t)$
\textit{will }display some oscillations. Their intensity depend on the
physical scales of the system. Similarly, we could evaluate the probability to
find the final state in the orthogonal combination leading to $w_{\nu_{\mu}%
}=\int d\mathbf{k}\left[  \left\vert -\sin\theta b_{1}(\mathbf{k}%
,t)+\cos\theta b_{2}(\mathbf{k},t)\right\vert ^{2}\right]  $. This expression
also leads to oscillations. But, if we do not distinguish among these two
configurations (because we do not measure neutrinos but only the mother and
the daughter states) we have to perform the sum and re-obtain the standard
formula: $w_{\nu_{e}}+w_{\nu_{\mu}}=w(t)=1-p(t)=1-e^{-\Gamma t}.$ Again, the
oscillations disappear. In the end, no matter how one designs the decay's
measurement of the mother and the daughter states: neutrino oscillations do
not generate time modulations.

\section{Non-exponential decay}

In QM the exponential decay is only an (extremely good) approximation:
deviations are predicted at both short and late times after the creation of
the unstable system \cite{facchiprl,duecan,ghirardi}. The very same effect has
been confirmed in QFT \cite{zenoqft}. Short-time deviations (related to a
residual correlation between unstable system and decay products) have been
clearly demonstrated in cold atoms experiments \cite{reizen}, while long-time
deviations were measured in molecular decays \cite{rothe}.

In Ref. \cite{gp}, we proposed that a similar deviation occurs in the
electron-capture decays of H-like ions due to the particular way in which the
energy eigenstate measurement is performed in a storage ring: no measurement
on the outgoing neutrino and long-lasting measurement of the daughter nucleus
\cite{gsi}. As a simple model, we consider the case in which the form factors
$f_{i}(k)$ generates a Breit-Wigner $d_{M}(E)=N\Gamma\left[  (E-M)^{2}%
+\Gamma^{2}/4\right]  ^{-1}$ which is cut by a cut-off $\Lambda$ in an energy
window centered a the value of the mass of the unstable system. The
corresponding survival probability
\begin{equation}
p(t)=\left\vert \int_{M_{M}-\Lambda}^{M_{M}+\Lambda}d_{M}(E)e^{-iEt}%
dE\right\vert ^{2}%
\end{equation}
shows an oscillating behavior with a period $T\sim1/\Lambda$ (see plots in
\cite{gp}).

In this interpretation, the physical origin of the cutoff is related (in a non
trivial way) to the time needed to measure the mass of the daughter nucleus,
which is of the order of 1sec. During this time interval the mother nucleus
and the decay products are still correlated and can therefore provide
significant deviations from the usual exponential decay law. As a consequence,
changing the detector would also affect the results (in particular, a detector
with a higher precision in the measurement of time would suppress the signal).

Within our model it is not possible to obtain oscillations in $\beta^{+}$
decays because the emitted positron is immediately adsorbed within the
detector and the correlation between mother and daughter states is broken
(similarly for electron capture decay experiments in which the nuclei are
embedded in a metallic matrix \cite{Vetter:2008ne}).

\section{Conclusions}

The not-yet clarified GSI oscillations need experimental
verification/falsification. Here, we have shown that neutrino oscillations
cannot be responsible for them, but that deviations from the exponential decay
law offer an interesting possibility. If the effect shall be confirmed, the
microphysics of the measurement process specifically used in the GSI
experiment needs to be theoretically modelled in a better way.

\end{document}